\documentstyle[12pt,psfig]{article}
\newcommand {\be}{\begin{equation}}
\newcommand {\ee}{\end{equation}}
\newcommand {\zm}{z_{\rm m}}
\newcommand {\zu}{z_{\rm unst}}

\begin{document}
%\title{Random Neighbor Theory of the Olami-Feder-Christensen 
%Earthquake Model}
%\author{Hans-Martin Br\"oker and Peter Grassberger\\
%Physics Department, University of Wuppertal\\
%42097 Wuppertal, Germany}
%\maketitle

\begin{center}
{\Large\bf Random Neighbor Theory of the Olami-Feder-Christensen 
Earthquake Model}\\[1 cm]

{\large\bf Hans-Martin Br\"oker$^1$ and Peter Grassberger$^{1,2}$}\\[0.5cm] 

$^1$ Physics Department, University of Wuppertal, D-42097 Wuppertal, Germany\\
$^2$ HLRZ c/o Forschungszentrum J\"ulich, D-52425 J\"ulich, Germany\\[0.5cm]
\bf{\today}
\end{center}

\begin{abstract}
We derive the exact equations of motion for the random neighbor version 
of the Olami-Feder-Christensen earthquake model in the infinite-size limit. 
We solve them numerically, and compare with simulations of the model for 
large numbers of sites. We find perfect agreement. But we do not find any 
scaling or phase transitions, except in the conservative limit. This is in 
contradiction to claims by Lise \& Jensen (Phys. Rev. Lett. {\bf 76}, 
2326 (1996)) based on approximate solutions of the same model. It
indicates again that scaling in the Olami-Feder-Christensen model is 
only due to partial synchronization driven by spatial inhomogeneities.
Finally, we point out that our method can be used also for other SOC 
models, and treat in detail the random neighbor version of the Feder-Feder 
model.
\end{abstract}
\newpage

\section{Introduction}
During the last ten years more than 2000 publications were concerned
with the idea of self--organized criticality (SOC) proposed by Bak, Tang
and Wiesenfeld(BTW)~\cite{bak}. They introduced a non--equilibrium system,
the so--called sandpile model, which is driven slowly by adding single 
sand grains at random positions. Without any control parameter to fine-tune, 
it evolves into a critical state. In this state the system reacts to the 
external drive with a series of relaxation events (avalanches). It becomes 
critical in the sense that the spatial and temporal distributions of these
avalanches obey power laws, indicating that any characteristic scales 
in space and time are lost. The attribute `self--organized' is to stress 
the absence of a fine-tuned control parameter.

A crucial point in understanding the robust scaling of the BTW model is 
the existence of a conservation law ~\cite{manna}: the total amount of 
sand in the system is 
conserved, if boundary effects and external perturbations are neglected.

In the frame of this concept, Olami, Feder, and Christensen (OFC) introduced
a nonconservative `continuous cellular automaton'~\cite{olami} as a specific
realization of the two--dimensional Burridge--Knopoff earthquake 
model~\cite{burr}. Details of this model will be described below. In 
contrast to the BTW model it is not conservative in general. It involves a 
parameter $\alpha$, and a conservation law holds only for a specific 
value $\alpha=\alpha_c$. It was found in ~\cite{olami,olami2} and subsequent 
simulations \cite{grass1} that the system displays power law 
behavior in a wide range of the control parameter $\alpha$ (not only near 
$\alpha_c$), and the critical exponents depend on $\alpha$. Thus the model 
seems to show SOC, and conservation seems not a necessary condition. 

But, on the other hand, it seems that spatial inhomogeneities are crucial for 
the observation of scaling in the OFC model \cite{grass1,corral,middleton}. 
In the original paper by OFC the boundary conditions (bc) were not periodic, 
which induced an inhomogeneity with a diverging length scale in the 
thermodynamic limit. This inhomogeneity of the bc leads to partial 
synchronization in the bulk which is both driven and destroyed by the 
boundary \cite{grass1}. Subsequent simulations with periodic bc showed no 
scaling \cite{socolar,grass1}, as did also simulations with frozen randomness 
but without diverging length scales \cite{jk,froyland}. The basic 
source of scaling in the OFC model is the slow build-up of large coherent 
domains in which the system itself is homogeneous, but which are driven by 
regions where the system is not homogeneous.

Although the definitions of these SOC models are simple, and they are easily 
simulated on a computer, only few exact results are known. Most of the 
difficulties in the analytical treatment arise from the spatial correlations 
due to the interactions of the particles. In a mean--field theory,
which is the first step towards a detailed understanding, these correlations
are simply neglected. A more refined strategy to avoid spatial correlations 
is to replace the nearest neighbors interactions by interactions between 
random sites. For the OFC model this was already attempted by~\cite{lise}. 
But in that paper additional assumptions and approximations were made which 
are hard to justify. With these assumptions, a transition was found 
from non--SOC to SOC at $\alpha$ significantly less than $\alpha_c$. 
This is very surprising, as we argued above that spatial structures 
are crucial for the emergence of scaling, and any such structures are of 
course eliminated in the random neighbor version.

In the following we study the random neighbor model in detail without any 
further approximations.
We will be led to a complete set of equations which allow us to calculate
numerically all the relevant quantities. We will see that there is no SOC
in the dissipative regime of the control--parameter. In the case of 
conservation the exact solution shows that the system becomes a critical 
branching process equivalent
to critical percolation on a Bethe tree, and the critical exponents take
their mean--field values.   

\section{The model}
The model lives on a set of $N$ sites, each of them equipped with 
a continuous stress (or `force') variable $z_i$. Each $z_i$ can take
any value $\geq 0$, but only values $<1$ are stable. After having initialized
each site with a randomly chosen value $z_i \in [0,1[$, the system evolves
according to the following rules:

(i) All $z_i$ are simultaneously and continuously increased with the
same speed $v=1$.

(ii) If any $z_i$ exceeds the threshold value $1$ the above driving stops,
and the forces are redistributed in the following way:\\
All unstable sites discharge simultaneously,
\be
   z_i \to 0 \qquad \forall z_i \geq 1\;.
                                                    \label{topprule1}
\ee
For each of these discharging sites $n$ random ``neighbors", 
$j_1, \ldots j_n $, are chosen and their
stress variables are increased by a fixed fraction of $z_i$.
\be
   z_{j_{k}} \to z_{j_{k}}+\alpha z_i \quad k=1,\ldots ,n
                                                     \label{topprule2}
\ee
The integer $n$ is constant but otherwise arbitrary.
If the application of eq.~(\ref{topprule2}) creates new unstable sites,
rule~(ii) is again applied in the next time step, again simultaneously for 
all unstable sites. This procedure is repeated until all sites are 
stable. After that, the system is again driven according to rule~(i), 
until at least one site with $z_i=1$ appears. A series of causally connected 
discharging events is called an earthquake or an avalanche. Its size $s$ 
is measured by the total number of discharges. If a site discharges $m$ 
times during an avalanche, it is counted $m$-fold in the calculation of $s$.
The duration $t$ of an earthquake is defined as the number of
sweeps through the lattice necessary to get a stable configuration. Obviously
$s$ as well as $t$ is always $\geq 1$.

The parameter $\alpha$ which controls the dissipation can take any value
between $0$ and $1/n$ ($\alpha>1/n$ is unphysical since sooner or later an 
infinite and ever growing avalanche 
would occur). Only for $\alpha = 1/n$ the system is conservative. Note 
that the randomness of the neighbors appearing in eq.~(\ref{topprule2}) is
annealed: For each discharging event, the $n$ random neighbors are chosen 
anew. Obviously this prevents that any spatial correlations in the values of
$z$ to build up. 

The numerical calculations as well as the simulations are restricted to the 
case $n=4$. Obviously this is most appropriate for a mean-field theory of 
the two-dimensional OFC--model. But our analytic results are more general 
and hold for any $n\geq 2$.

As well known, the original (nearest neighbor) version of the model is 
very sensitive to the choice of bc's \cite{socolar,grass1}. Any bc 
other than periodic introduce inhomogeneities which are crucial in 
building up the spatial structures which manifest themselves in non-trivial 
avalanches \cite{grass1,corral,middleton}. For the random neighbor version, 
non-periodic bc were used in \cite{lise}. This also introduces spatial 
inhomogeneity which is however completely irrelevant for the dynamics,
`space' being a dummy concept in a random neighbor model. In addition, the 
bc used in \cite{lise} lead to specific finite size corrections which 
might be not easy to disentangle from the true asymptotic behavior. 
In contrast, we treat all sites equally in the present paper,
mimicking thereby periodic bc. In addition, we shall only study the 
infinite size limit. More precisely, we shall formally work with a 
finite number $N$ of sites, but will understand that we are only 
interested in the limit $N\to\infty$. For finite sizes there are 
correlations which make the study of the model rather awkward.

\section{Random Neighbor Theory} 

In the OFC model, there is a finite chance that two sites become unstable 
simultaneously during the continuous increase (i). It arises from the 
non-zero probability that two sites which had discharged in the same 
previous earth quake have not been hit by a discharging neighbor (or have 
been hit by the same neighbors) until 
they reach $z=1$. In the lattice version this implies that the notion 
of an earth quake itself becomes a bit delicate: should we consider an 
event which was triggered simultaneously by two sites as one 
earth quake or two? In the off-lattice version we still have a non-zero 
chance for such events. But on an infinite lattice the sub-quakes following 
each unstable site will not overlap. Thus they will evolve completely 
independently. This means that the model becomes effectively abelian 
\cite{dhar} in the sense that we can change the order of updates in 
different sub-events. Also, we can associate earth quakes uniquely with 
the original unstable sites which triggered them. In the following, we 
will always define earth quakes in this way. An event which started 
with $k$ sites becoming unstable is counted as $k$ earth quakes, separated 
by infinitesimal time delays and taking place in arbitrary order.

So we can assume without loss of generality that after the relaxation of 
an earthquake there is exactly one site which has a stress value greater
than all the others, and which will be the seed of the next avalanche.
The value of this stress immediately after the earthquake has stopped 
will be called $\zm$. Its mean value, averaged over all
earthquakes, is denoted by $\overline{\zm}$. Since we consider 
the large system limit, $\zm$ will not be correlated with the 
size of the previous avalanche. This is our crucial assumption, and it 
depends on the fact that we can neglect `global' avalanches whose 
size is comparable to the total size of the system. In this limit the 
model thus becomes a branching process with time dependent branching 
rates. We shall later verify that this assumption is self consistent, 
and is true in simulations.

The average increase of
the force on each of the $N$ sites between two earth quakes due to the 
external driving is then given by $1-\overline{\zm}$. On the other
hand, each discharge dissipates an average value of $(1-n\alpha ) 
\overline{\zu}$, where $\overline{\zu}$ is the mean force
on the unstable sites, averaged over all discharging events.
In the stationary state, when $\sum z_i$ fluctuates around a
constant value, the external increase must be exactly compensated
by the average dissipation. This gives an exact formula involving the 
average earthquake size $\langle s\rangle$ defined as the mean number 
of discharges per earth quake,
\be
   (1-n\alpha)\;\overline{\zu}\;\langle s\rangle = N(1-\overline{\zm})\;.
                                         \label{emitt.gl}
\ee
Notice that the product of averages on the l.h.s. does not result from 
a factorization approximation but from the definition of $\overline{\zu}$, 
and is exact. Therefore, 
this equation is correct even if the above mentioned simplifying 
assumptions are not true, and holds thus also in the fixed neighbor 
version of the model.
Since the left hand side of this equation remains finite for $N\to\infty$
(as long as $\alpha <1/n $), we see that $1-\overline{\zm}\propto 1/N$.

On the other hand, since the force increase between earth quakes is 
assumed to be with velocity $v=1$, the average time between two earth
quakes is given by $1-\overline{\zm}$. On a `macroscopic' time scale 
where we neglect the duration of earth quakes compared to the inter-quake 
times (this assumption is inherent in the model), the {\it toppling 
rate} is thus given by 
\be 
   \sigma = \frac{\langle s \rangle}{N(1-\overline{\zm})}
	= \frac{1}{(1-n\alpha)\overline{\zu}} \;.
                                                \label{sigma.gl}
\ee
This tells us how frequently each site discharges per time unit. The 
rate to {\it be hit} according to eq.(\ref{topprule2}) is then given by
$n\sigma$. 

Let $P(z)$ be the probability density for a given site to have a force 
value $z$ (from now on, we shall consider only $N=\infty$). Obviously, 
$P(1)$ is the rate with which new earth quakes are initiated, while $P(0) 
=\sigma$ is the rate with which new force-free sites are created by 
discharges. Therefore, 
\be
   \langle s\rangle = P(0) / P(1) \;.           \label{s-av} 
\ee
Similarly, $P_j(z)$ denotes the joint probability density that a site has 
a value $z$ and was hit exactly $j$-times since its last discharge. Since 
we consider only $N=\infty$ and have argued that global avalanches are 
negligible in this limit, $P(z)$ and $P_j(z)$ do not fluctuate with time.
Obviously, we have
\be
   P(z) = \sum_{j=0}^m P_j(z) \qquad z \in [0,1]\;.
\label{pdens.gl}
\ee
Because a hit increases $z$ at least by an amount $\alpha $, each $P_j(z)$ 
vanishes exactly for $z < j\alpha $. Therefore the upper limit $m$ in the 
above sum is given by the largest integer for which $m\alpha \leq 1$.
For later use we define the integrated distribution as
\be
{\cal P}(z) = \int_z^1 P(z') dz'\;.            \label{calP}
\ee

To obtain $P_0(z)$ we notice that the probability to be hit exactly
$k$ times during a time interval $z$, when the rate is $n\sigma$, 
is given by the Poisson distribution
\be
   \pi_k(z) =\frac{1}{k!} (n\sigma z)^k\; e^{-n\sigma z}\;. 
\ee
This leads to 
\be 
   P_0(z)=\sigma \pi_0(z)= \sigma e^{-n\sigma z}.
\ee

The other $P_j(z)$ depend on the distribution of the amount $\Delta z$ 
which a site receives when it gets hit by a discharge. This in turn 
depends on the distribution of forces of unstable sites at the moment 
of their discharge. We denote the density of this distribution by $C(z)$.
It is related to $\overline{\zu}$ by
\be 
   \overline{\zu} =\int_1^{\infty} z\;C(z)dz.
\label{zcrit.gl}
\ee
(Here and in the following, integrals over functions with $\delta$-peaks 
at the integration limits are understood as containing all contributions 
from these peaks, $\int_a^b f(x)dx \equiv \lim_{\epsilon\to 0}
\int_{a-\epsilon}^{b+\epsilon} f(x)dx$.)
The first site of any earthquake discharges exactly with $z=1$. This 
gives a delta contribution to $C(z)$, with relative weight 
$1/\langle s \rangle$. We can therefore make the ansatz
\be
   C(z)= \frac{1}{\langle s \rangle} \delta (z-1) + \tilde{C}(z).
%       (1-\frac{1}{\langle s \rangle}) \tilde{C}(z).
                                                     \label{czerl.gl}
\ee
The second term corresponds to all subsequent discharges. About the 
function $\tilde{C}(z) $ we know that it has to vanish for all $z$
outside the interval $[1,1/(1-\alpha )]$. The upper limit would be
reached if an infinite earthquake contained a series of successive
hits onto sites with maximum value $z=1$. This upper limit could be 
surpassed only if a site were hit simultaneously by two discharges, but 
the chance for this is zero on an infinite lattice. The amount of force
$\Delta z$ that a discharging site drops onto each of the $n$ random 
neighbors is then distributed according to
\be
   Q_1(\Delta z)=\alpha^{-1}C(\Delta z/\alpha)\;,\qquad 
    {\rm supp}\; Q_1 = [\alpha,\alpha/(1-\alpha)]\; . 
                                          \label{qvonc.gl}
\ee
Similar to eq.~(\ref{czerl.gl}) we can write $Q_1(\Delta z)$ as
\be
   Q_1(\Delta z)= \frac{1}{\langle s \rangle} \delta (\Delta z-\alpha) + 
       \tilde{Q}_1(\Delta z).
                                                   \label{qzerl.gl}
\ee
The convolution integrals
\be
   Q_k(\Delta z)= \int_{\alpha}^{\Delta z} 
	Q_{k-1}(\Delta z-\Delta z')Q_1(\Delta z')d\Delta z', 
                   \quad k \geq 2
                                                  \label{qconv.gl}
\ee
then give us the probability densities for the total increase of force 
when a site was hit exactly $k$ times. Note that $Q_k(\Delta z)$ vanishes
for $\Delta z$ outside the interval $[k\alpha, k\alpha /(1-\alpha)]$.
We see finally that every $P_j(z)$ has to obey
\begin{eqnarray}
P_j(z) &=& P_0(0) \int_{j\alpha}^{z} \pi_j(z-\Delta z) Q_j(\Delta z) d\Delta z 
	\nonumber \\
       &=& \frac{\sigma}{j!} \int_{j\alpha}^{z} \left(n\sigma(z-\Delta z)
           \right)^j e^{-n\sigma (z-\Delta z)} Q_j(\Delta z) d\Delta z.
\label{pjconv.gl}
\end{eqnarray}  

Let us now come back to the function $\tilde C(z)$ which describes the 
distribution of sites which have become unstable by being hit by a 
discharge. It is obtained by the convolution of the distribution of 
(stable) sites before they are hit, with the distribution of amounts 
received during the destabilizing discharge, 
\be
   \tilde C(z) = n \;\Theta(z-1)\; \int_{\alpha}^{\alpha/(1-\alpha)} 
         P(z-\Delta z) Q_1(\Delta z) d\Delta z
                                                  \label{csch.gl}
\ee
Here $\Theta(x)$ is the Heaviside step function. It takes into account 
that $\tilde C(z)$ is supposed to describe only those sites which 
actually do become unstable. The factor $n$ takes into account that 
each discharge event --- the probability density of which is given by 
the integrand --- gives rise to $n$ potentially unstable sites. The 
integration limits are given by the support of $Q_1$, 
see eq.(\ref{qvonc.gl}).

Let us now consider the integral of $\tilde C(z)$ over all $z$. 
Interchanging the integrations we obtain
\be
   \int_1^\infty \tilde C(z)\;dz = n \int_{\alpha}^{\alpha/(1-\alpha)} 
       d\Delta z\; Q_1(\Delta z) \int_1^\infty P(z-\Delta z)\; dz \;.
\ee
The inner integral on the rhs. is just ${\cal P}(1-\Delta z)$ (see 
eq.(\ref{calP})), while the left hand side is equal to
$ 1-1/\langle s \rangle $ due to eq.(\ref{czerl.gl}). 
Rearranging terms we arrive thus at a second equation for 
$\langle s\rangle$, 
\be
   \langle s\rangle = \left[1-n\int_{\alpha}^{\alpha/(1-\alpha)} 
        {\cal P}(1-z) Q_1(z) dz\right]^{-1}\;.
                                                \label{emvps.gl}
\ee

We claim that the above equations are complete in the sense that they 
fix the solution uniquely, for each $\alpha < 1/n$. To show this, and 
to provide also a practical method to solve them numerically for not 
too large $\alpha$, we give a recursion 
scheme which converges to the solution as the iteration level $r$ 
tends to infinity, at least for sufficiently small $\alpha$. For 
larger values of $\alpha$ the recursion might not be practical, but the 
set of equations should still fix the solution by continuity. Notice 
that different recursion schemes are in principle possible where the 
order of replacements is changed in various places. 

To start the recursion, we select a desired accuracy $\eta$ and 
choose the initial distribution 
$Q_1(z)^{(0)}$ in some arbitrary way. It need not even be normalized. 
For small $\alpha$, a good choice is $Q_1(z)^{(0)}$ constant. For 
$\alpha \approx 1/n$ we can also take $Q_1(z)^{(0)} = \delta(z-1/n)$. 
In the recursive step from $r-1$ to $r$ we do the following:\\
(1) (re-)normalize $Q_1(z)= const \times Q_1^{(r-1)}(z)$ with
$const = [\int Q_1^{(r-1)}(z)dz]^{-1}$; \\
(2) compute $\sigma$ from eqs.(\ref{sigma.gl}), (\ref{zcrit.gl}), and 
(\ref{qvonc.gl}), 
\be
   \sigma = \alpha\left[(1-n\alpha)\;\int_\alpha^{\alpha/(1-\alpha)} 
       Q_1(z)\;z\;dz\right]^{-1} \; ;
\ee
(3) compute $Q_k(z)$ for $k>1$ by means of eq.(\ref{qconv.gl}); \\
(4) compute $P_0(z)=\sigma e^{-n\sigma z}$, compute $P_k(z)$ for 
$k>0$ by means of eq.(\ref{pjconv.gl}), and obtain $P(z)$ as 
$\sum_k P_k(z)$;\\
(5) compute the new $\langle s\rangle$ from eq.(\ref{s-av});\\
(6) compute the new $Q_1(z)$ from eqs.(\ref{qvonc.gl}), 
(\ref{czerl.gl}), and (\ref{csch.gl}),
\be
   Q_1^{(r)}(z)  =  {1\over \langle s\rangle} \delta(z-\alpha) + 
        {n\over \alpha}\;\Theta(z-\alpha)\; 
        \int_{\alpha}^{\alpha/(1-\alpha)} P(z/\alpha-\zeta) Q_1(\zeta) 
          d\zeta \; ;
\ee
(7) if $\sigma$ or $\langle s \rangle$ have changed by a fraction larger 
than $\eta$, then goto (1);\\
(8) verify that the normalization constant in step (1) is unity within 
some acceptable error, and that $\langle s \rangle$ satisfies 
eq.(\ref{emvps.gl}).

We have not shown formally that this iteration converges always, but we 
have done extensive numerical investigations. The scheme converges very 
fast and stably for small $\alpha$, but convergence is slowed down 
when $\alpha \to 1/n$. For this reason we had problems to obtain 
solutions for $\alpha\geq 0.24$, although the recursions shows no sign 
for divergence even in such extreme cases. Also, numerical errors in the 
integration routines tend to accumulate for $\alpha \to 1/n$, rendering 
in particular the estimate of $\langle s\rangle$ problematic.
Since the integrands are not analytic functions, it 
does not make sense to use very sophisticated integration routines. We 
used the extended trapezoidal rule with up to $10^4$ points. 

\begin{figure}[ht]
\centerline{\psfig{file=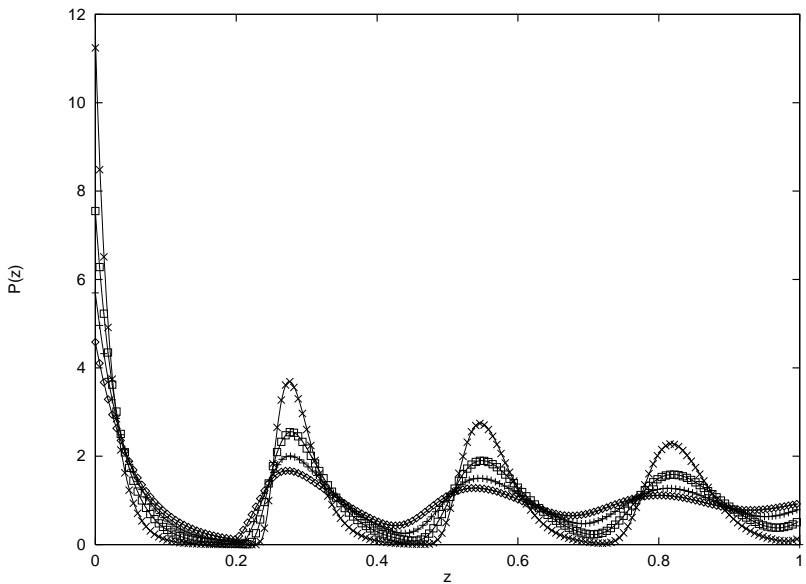,width=11cm,angle=270}}
\caption{
\small Probability density \protect{$P(z)$} against $z$. The continuous 
lines are the predictions from the theory and the points show the results 
obtained from simulations. They fall perfectly on top of each other on 
the scale of this figure. The four curves are for \protect{$\alpha $} 
= 0.20, 0.21, 0.22, and 0.23, in order of increasing sharpness of 
the peaks.
} 
\label{rho.fig}
\end{figure}

Results for 
$n=4$ are shown in fig.~\ref{rho.fig}, where we also compare with 
straightforward simulations of eqs.(\ref{topprule1}),(\ref{topprule2}). 
For the latter we typically used $N=10^6$ to $8\times 10^6$, and discarded 
transients of up to $2\times 10^6$ iterations. No difference between 
theory and simulation is detectable in fig.~\ref{rho.fig}. This shows 
that the numerical integration was sufficiently accurate, the iterations 
had converged, $N$ was sufficiently large to have negligible finite 
size corrections, and the discarded transients were sufficiently long. 
Qualitatively, fig.~\ref{rho.fig} is similar to 
fig.1 in \cite{lise}, but the first peak in that paper seems much too 
high. It is not clear whether this results from the bc used in that 
paper or from transients. For $\alpha=0.23$, we find $P(0)\approx 11.5$ 
both from simulations and from the analytic solution, while $P(0)\approx 
33$ is quoted in \cite{lise}.

\begin{figure}[ht]
\centerline{\psfig{file=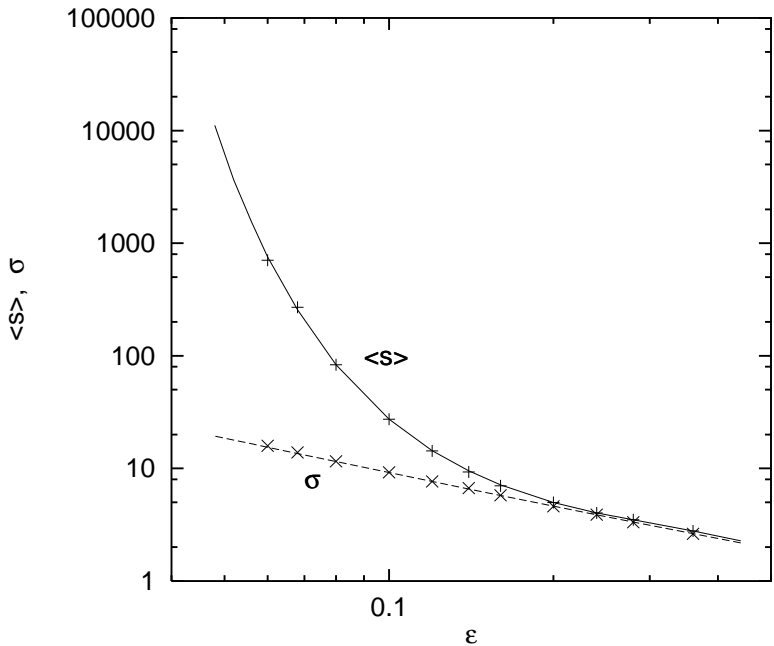,width=11cm,angle=270}}
\caption{\small Log-log plot of $\sigma$ and $\langle s\rangle$ against 
$\epsilon = 1-4\alpha$. Continuous lines are from theory, points from 
simulations. }
\label{sigs.fig}
\end{figure}

\begin{figure}[ht]
\centerline{\psfig{file=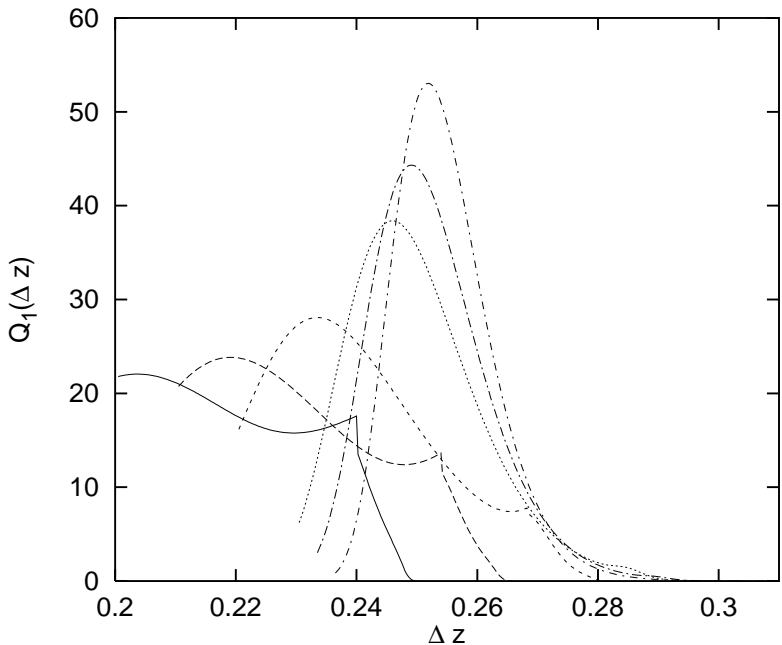,width=11cm,angle=270}}
\caption{\small \protect{$\tilde{Q}_1(\Delta z)$} against $\Delta z$, 
for the same 
values of \protect{$\alpha $} as in fig.~\ref{rho.fig}, and in addition 
for \protect{$\alpha = 0.233$} and \protect{$\alpha = 0.236$}. For 
clarity only theoretical predictions are shown. The cusps are at $\Delta z
=\alpha+\alpha^2$ and correspond to the maximal $\Delta z$ transfered by 
first generation descendents of the avalanche seed.
}
\label{Q.fig}
\end{figure}

In fig.~\ref{sigs.fig} we show $\sigma$ and $\langle s\rangle$ as functions 
of $\epsilon = 1-4\alpha$. We see that $\sigma \approx 1/\epsilon$, while 
$\langle s\rangle$ diverges much faster when $\alpha \to 1/4$. Finally, in 
fig.~\ref{Q.fig} we show $\tilde{Q}_1(z)$. This shows a very interesting 
qualitative change as $\alpha$ approaches 1/4. For $\alpha < 0.23$, 
$\tilde{Q}_1(z)$ is centered at $z<1/4$. Its center moves to the right 
as $\alpha$ increases, reaching a value slightly larger than 1/4 for 
$\alpha \approx 0.233$. After that, its center moves very little, and 
it just shrinks slowly to a $\delta$-function centered at 1/4.

The most important result is that we see no hint of any singularity for 
$\alpha < 1/4$, as predicted in \cite{lise}, and we also see no mechanism 
which could lead to such a singularity. Indeed, we can prove rigorously 
that $\langle s\rangle < \infty$ for all $\alpha <1/n$. This follows 
simply from the fact that $\sigma \leq 1/(1-n\alpha)$ due to eq.(4), and 
$\langle s\rangle =\sigma/P(1)\leq \sigma/P_0(1)=e^{n\sigma}$ due to 
eqs.(5) and (9). Conversely, this argument shows that $P(1)$ must tend to 
0 for $\epsilon\to 0$, as also shown by the numerics. 

According to \cite{lise}, a singularity with 
$\langle s\rangle \to\infty$ should occur for $n=4$ at $\alpha = 2/9 = 
0.222\ldots.$ We believe that this is due to unjustified assumptions made 
in \cite{lise}. Another important result is that $P(z)$ is finite and 
non-zero at $z=1$ and at $z=0$. This shows that `global' earth quakes 
have indeed no effect, as they would lead to a depletion at $P(z)$ at 
$z=1$ or an infinity at $z=0$ due to eq.(\ref{s-av}).

\section{The Limit \boldmath $\alpha \to 1/n$\unboldmath}

Figure 1 suggests that $P(z)$ tends to a sum of four delta peaks
at multiples of $\alpha$, for $\alpha \to 1/4$. More generally, we 
expect $P(z)$ to tend towards a sum of $n$ delta peaks at $z=k/n$, 
$k=0,\ldots n-1$ (this is reminiscent of a generalized sandpile model 
with real-valued heights by Zhang \cite{zhang}). 
We shall see that this is indeed a valid solution 
after proper rescalings of $\sigma $ and $\langle s\rangle $. 

Formally we introduce 
\be
   \epsilon = 1-\alpha n \;,
\ee
and consider the limit $\epsilon\to 0$. We shall argue that a self 
consistent solution for $P(z)$ in this limit is a sum of delta peaks. 
If this is true, only sites which 
have already $z\approx (n-1)/n$ will become unstable by receiving an 
extra $\Delta z\approx 1/n$, and hence $\zu \to 1$ for $\epsilon\to 0$.
From this we see on the one hand that $\sigma = 1/\epsilon$ to leading 
order in $1/\epsilon$, which in turn gives $\sigma \pi_k (z) \to 
n^{-1}\delta(z)$ for each value of $k$. On the other hand it gives 
$C(z) = \delta(z-1)$ and $Q_1(\Delta z)= \delta (\Delta z-1/n)$. The 
latter implies $Q_k(\Delta z)= \delta (\Delta z-k/n)$ for any $k\geq 2$, 
which finally gives 
\be
   P(z)= \frac{1}{n} \sum_{j=0}^{n-1} \delta(z-j/n),
                                                \label{pcrit.gl}
\ee
i.e. our initial assumption was self consistent.

In spite of the simplicity of this solution, we should be careful 
in interpreting it, as several limits are involved. The easiest way is
to take first the infinite volume limit, and then $\epsilon\to 0$. If 
we want to take the limit $\epsilon\to 0$ first, we have to use 
absorbing sites which mimic absorbing bc's, but it is not a priori 
clear how their number should scale with $N$ (for a related problem 
in a mean field version of the abelian sandpile model, see 
\cite{flyv,bundschuh}).

While the behavior exactly at $\alpha=1/n$ is thus well understood, 
we were not able to find an analytic solution for finite $\epsilon$. 
But we can give approximate solutions for small $\epsilon$, and predict 
the behavior of $\langle s\rangle$ for $\epsilon\to 0$. 
For small but finite $\epsilon$, we approximate each $Q_j(z)$ by a 
delta function at $z=j/n$. Then $P_j(z)$ is roughly given by 
\be
   P_j(z) \approx \Theta(z-j/n) \;
       {\sigma\over j!} [(z-j/n)\sigma]^j e^{-(z-j/n)\sigma}\;.
\ee
From this, eq.(\ref{s-av}), and the fact that $P(0)=\sigma$, we get 
\be
   \langle s \rangle = \sigma /P(1) \approx \sigma / P_{n-1}(1) 
       \approx {(n-1)! \over \sigma^{n-1}} e^\sigma
       \approx (n-1)! \epsilon^{n-1} e^{1/\epsilon} \;. 
                                      \label{s-eps}
\ee
There are substantial corrections to this, mainly from the contribution 
of $P_n(1)$ to $P(1)$, which are hard to estimate. Thus,  
the actual values of $\langle s \rangle$ are smaller than given 
by eq.(\ref{s-eps}), but 
eq.(\ref{s-eps}) gives the correct trend. In particular, it explains 
why $\langle s \rangle$ diverges extremely fast for $\alpha \to 1/n$, 
making simulations in this limit very difficult.

\section{Earth Quake Statistics}

In the previous sections we have only studied the force distribution 
and the average earth quake size. In order to discuss the distribution 
of earth quake sizes and durations, we need some more definitions 
and some basic results from the theory
of branching processes as found e.g. in \cite{har}.

For any integer $i\geq 1$ we define $p_i$ as the probability that a site 
becomes unstable if it is hit by a discharge event during the $(i-1)$-st 
generation of an earth quake, and will therefore discharge itself in 
the $i$-th generation. For the first generation, $p_1$ is the 
probability that a site which receives $\Delta z =\alpha $ gets unstable. 
Thus, simply $p_1={\cal P}(1-\alpha)$. 
For general $\alpha$, the other probabilities $p_i$ depend on the height 
distribution of the unstable sites in the previous generation. But 
for $\epsilon\to 0$ all discharging sites have $z=1$, and $p_i$ is 
simply the probability that the hit site is in the $(n-1)$-st peak, 
$p_i = p=1/n $ for all $i>0$. 

In the following we shall therefore discuss only the case $\epsilon=0$, 
deferring the general case to the end of this section.

We assume thus that $p_i = p=1/n $ for all $i>0$. The
probability that an unstable site creates $l$ unstable offsprings in the
next generation is then given by
\be 
w_l = {n \choose l} (1-p)^{n-l} p^l 
\ee
 with the generating function
\be
g(u)=\sum_{l=0}^n u^l w_l = (1-p+up)^n.
\label{gene.gl}
\ee
With the use of $g(u)$ it
becomes very easy to calculate recursively the distributions of the size 
and the duration of the earthquakes. In order to do this we introduce
the iterates 
\begin{eqnarray}
g_0(u)&=& u, \quad g_1(u) = g(u), \nonumber \\
g_{m+1}(u) &=& g[g_m(u)], \quad m=1,2,\ldots , \;.
\end{eqnarray}
The probability that the earthquake stops in the $t$--th time step is given 
by~\cite{har}
\be
{\cal P}_t= g_t(0)-g_{t-1}(0) \quad \mbox{with} \quad t=1,2,\ldots .
\ee

\begin{figure}[ht]
\centerline{\psfig{file=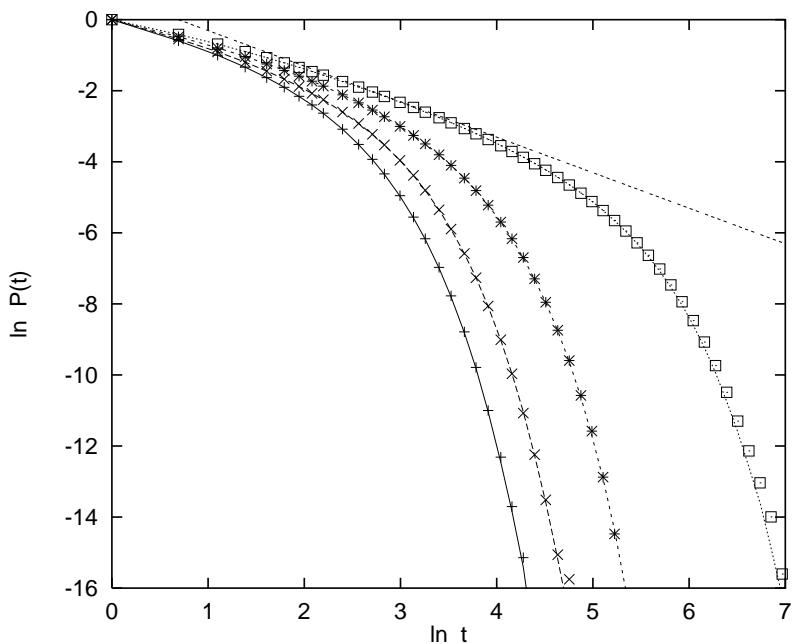,width=11cm,angle=270}}
\caption{\small The integrated distribution \protect{$P_t$}. Again, lines 
show the theoretical predictions and points are from simulations. Increasing 
from left to right, \protect{$\alpha $} takes the same values as in fig.~1. 
The dashed line shows eq.\ (\protect{\ref{ptcrit.gl}}).} 
\label{tdis.fig}
\end{figure}

The integrated distribution
\be
P_t =\sum_{t'=t}^{\infty} {\cal P}_{t'} =1-g_{t-1}(0)
\label{ptrec.gl}
\ee
denotes the probability that an avalanche lasts for $\geq t$ time steps.
Eq.\ (\ref{ptrec.gl}) can be used directly to calculate $P_t$ for small $t$, 
whereas for large $t$ we use the asymptotic behavior. The chain of 
identities
\begin{eqnarray}
P_{t+1} &=& 1- g_t(0) \nonumber \\
	&=& 1- g(g_{t-1}(0)) \nonumber \\
	&=& 1- g(1-P_t) \nonumber \\
	&=& 1- (1- pP_t)^n
\end{eqnarray}
leads to
\be
\frac{d}{dt}P_t \simeq P_{t+1}-P_t = -{n\choose 2} p^2 P_t^2 + o(P_t^3)
\ee
with the solution
\be
P_t \sim \frac{2n}{n-1} t^{-1}\;.
\label{ptcrit.gl}
\ee
This is a special case of the general theorem~\cite{har}
\be
P_t \sim \frac{2}{tg''(1)}
\ee
for a critical branching process.

The next quantity of interest is the size distribution of the earthquakes.
With ${\cal D}_s $ we denote the probability that the size of an avalanche
is exactly $s$. While ${\cal D}_1 = (1-1/n)^n $ is obvious, the calculation 
of ${\cal D}_s $ for $s>1$ proceeds as in~\cite{broek}. We first denote 
by $a_k^{(s)}$ the $k$--th Taylor coefficient of $[g(u)]^s$, i.e. 
$[g(u)]^s = a_0^{(s)}+ a_1^{(s)}u + a_2^{(s)}u^2 + \ldots$ .
A theorem due to Dwass~\cite{dwass} tells us then that
\be
   {\cal D}_s = \frac{1}{s} a^{(s)}_{s-1}, \quad s \geq 1\;.
\ee
In the present case, we have 
\be
   a_k^{(s)}= {ns \choose k} (1-p)^{ns-k}p^k\;,
\ee
leading to
\be
   {\cal D}_s =\frac{1}{s}{ns \choose s-1} (1-p)^{(n-1)s+1}p^{s-1}.
\ee
The local limit theorem of Moivre--Laplace states that in the limit
$s \to \infty $ the distribution ${\cal D}_s$ with $p=1/n$ tends to
\be
{\cal D}_s \approx  \frac{1}{\sqrt{2\pi(1-1/n)}} s^{-3/2}.
\ee
This means that the system gets critical for $\alpha =1/n$, and the 
critical exponents take the same values as for mean--field percolation
\cite{stauffer}.

\begin{figure}[ht]
\centerline{\psfig{file=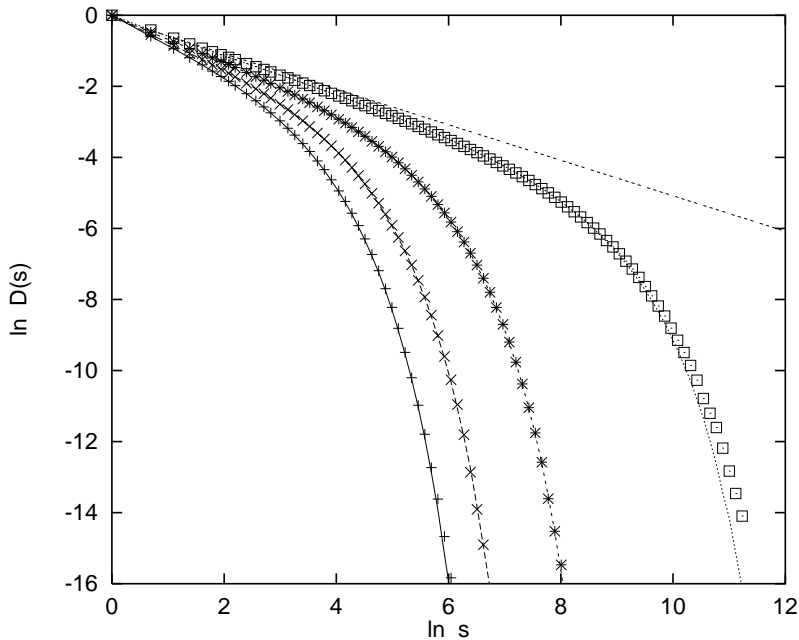,width=11cm,angle=270}}
\caption{\small The integrated distribution 
\protect{$D_s= \sum_{s'=s}^{\infty} {\cal D}_{s'}$}
with the same \protect{$\alpha$--values as in fig.\ \ref{tdis.fig}}. The 
dashed line shows the scaling law 
\protect{$D_s= \frac{2}{\sqrt{2\pi(1-1/n)}} s^{-1/2} $}.
} 
\label{sdis.fig}
\end{figure}

In the subcritical phase the arguments are more tedious. Let us denote by 
$c_i(z)$ the joint probability distribution that a discharge happens 
during the $i$-th generation of an earth quake, and that the discharging 
site has force $z$. It is related to $C(z)$ and to $p_i$ by 
\be
   C(z) = \frac{1}{\langle s\rangle}\delta(z-1)+ c_1(z) +
             c_2(z) +\ldots.
\ee
and 
\be
   p_i = \frac{\int_{1}^{1/(1-\alpha)} c_i(z)\; dz}
              {n\int_{1}^{1/(1-\alpha)} c_{i-1}(z)\; dz} \;.
\ee
This is easily checked by noting that it is compatible with 
\be
   \langle s\rangle = 1+np_1+n^2p_1p_2+n^3p_1p_2p_3+\ldots \;.
\ee
The functions $c_i(z)$ satisfy a recursion relation similar to 
eq.(\ref{csch.gl}), 
\be
   c_i(z) = n\; \Theta (z-1)\; \int_1^{1/(1-\alpha)}
         P(z-\alpha z')\; c_{i-1}(z')\; dz'
\ee
with $c_0(z)=\delta(z-1)/\langle s\rangle$.
Again we were not able to solve this analytically. But 
given a numerical estimate of $P(z)$ we can solve it numerically for 
$c_i(z),\;i=1,2,\ldots$, from which we obtain $p_i$ by integration.
Again this was done only for $n=4$. For each considered value of $\alpha$
we found that $p_i$ increases monotonically with $i$ and converges very
quickly to a constant value $< 1/4$. This is easy to understand. The 
increase is due to the fact that the first discharging site has $z=1$, 
while all subsequent ones have $z$ slightly larger than 1. The fact that 
$p_i<1/4$ reflects the fact that we are dealing with a subcritical 
branching process. 

In contrast to the critical case we now have a time dependent 
(non-autonomous) branching process, i.e.\ the
generating function $g(u)$ depends on the generation. The
mathematical treatment becomes now more tedious. For the further comparison 
between simulations and theory we therefore used the theoretically obtained 
$p_i$ to simulate a branching process, and compared the results with 
direct simulations of the OFC model. Figures \ref{tdis.fig} and 
\ref{sdis.fig} show that the agreement is essentially perfect, except for 
large $s$ and $t$, and for $\alpha=0.23$. The discrepancies seen there 
arise from the numerical problems mentioned in sec.3.

\section{The Feder-Feder Model}

Up to now we have considered only the OFC model, but our methods are 
much more general. To illustrate this, we shall discuss in this section 
the random-neighbor version of a model introduced by Feder and Feder (FF)
\cite{feder}. 
The FF model is identical to OFC model, with the only exception that
eq.(\ref{topprule2}) is replaced by
\begin{equation}
z_{j_k} \to z_{j_k} +\alpha\;, \qquad k=1,\ldots,n.
\end{equation}
This means that a site hit by a discharge always receives a fixed amount 
$\alpha $ regardless of the $z$--value of the discharging site. This leads 
to a significant simplification of the equations of motion. To derive
them, we have first of all to notice that the {\it toppling rate} is now
given by
\begin{equation}
\sigma = \frac{\langle s\rangle}{N(1-\overline{z_m})}
       = \frac{1}{\overline{z_{unst}}-n\alpha}.
\end{equation}
Since $z_{unst} \geq 1$, it is not a priori clear whether $\sigma $
diverges in the limit $\alpha \to 1/n $. But we will
show that this is indeed the case.

The main simplification arises from the fact that $Q_1(\Delta z) $ 
uncouples from $C(z)$ and is given by
\be
   Q_1(\Delta z) = \delta(\Delta z - \alpha) \;.
\ee
From this one obtains immediately
\be
   Q_k(\Delta z) = \delta(\Delta z - k\alpha)\;,
\ee
see eq.(\ref{qconv.gl}), and
\be
P(z) =\sum_{j=0}^m P_j(z) 
     =\sum_{j=0}^m \frac{\sigma}{j!}(n\sigma(z-j\alpha))^j 
	\Theta(z-j\alpha) e^{-n\sigma(z-j\alpha)},
\ee
where $m$ is again the largest integer $\leq 1/\alpha$. To obtain 
$\sigma$ as a function of $\alpha$, we use the normalization condition
$\int_0^1 P(z) dz = 1$ which gives
\be
   n = \sum_{j=0}^m{1\over j!}\int_0^{(1-\alpha j)n\sigma} 
               dx\;x^j\;e^{-x} \;.             \label{n-ff}
\ee
For $\alpha$ close to $1/n$ we have $m=n$, and this condition can be 
rewritten as
\be
   \int_0^{\epsilon n\sigma} dx\;x^n\;e^{-x} \;=\;
     \sum_{j=0}^{n-1}{n!\over j!}\int_{(1-\alpha j)n\sigma}^\infty 
              dx\;x^j\;e^{-x} \;,
\ee
where we have used again $\epsilon = 1-n\alpha$. From this it is easily 
seen that $\sigma\to\infty$ for $\epsilon\to 0$. Otherwise, the left 
hand side would tend to zero in this limit, while the r.h.s. would 
remain non-zero. But if $\sigma$ diverges, the r.h.s. is dominated by 
the term with $j=n-1$. Keeping only dominant terms we arrive at
\be
   \sigma \approx (n+1) \log(1/\epsilon) \;.
\ee

The mean avalanche size can again be calculated from eq.(\ref{s-av}),
\be
   \langle s\rangle = \sigma/P(1) 
                    = \left[\sum_{j=0}^m \pi_j(1-j\alpha)\right]^{-1}.
                                     \label{s-ff}
\ee
Obviously $\langle s\rangle$ is finite for all $\alpha<1/n$ and diverges 
for $\alpha\to 1/n$. In this limit the sum is dominated by the term with 
$j=n$, giving
\be
   \langle s\rangle \approx n!\;(n\sigma\epsilon)^{-n} 
        \sim \epsilon^{-n}\;,
\ee
up to constant and logarithmic factors in $\epsilon$ which could easily 
be computed.

Thus $\sigma$ and $\langle s\rangle$ both diverge much slower than in 
the OFC model. This reflects the increased dissipation in the FF model. 
Exact values of $\sigma$ and $\langle s\rangle$ obtained numerically 
from eqs.(\ref{n-ff}) and (\ref{s-ff}) are shown in fig.6. For small 
$\epsilon$ one finds good agreement with the asymptotic predictions.
Avalanche dynamics can be treated exactly as in the OFC model.

\begin{figure}[ht]
\centerline{\psfig{file=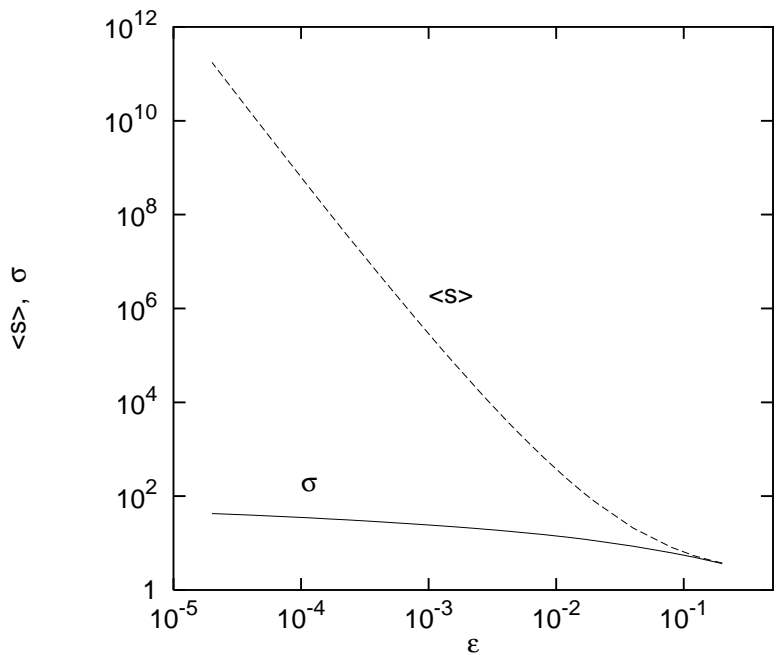,width=11cm,angle=270}}
\caption{\small Log-log plot of $\sigma$ and $\langle s\rangle$ against 
$\epsilon = 1-4\alpha$ for the Feder-Feder model. }
\label{ff-sigs.fig}
\end{figure}

\section{Conclusion}

Our results show clearly that there are neither scaling nor phase 
transitions in the random neighbor version of the dissipative 
Olami-Feder-Christensen earthquake model. Scaling is observed only in the 
conservative limit, in which case one has a critical branching process. 
This is in direct contradiction to claims in \cite{lise}. 
The latter was based on approximate random neighbor equations, while the 
present work is based on the exact equations. These equations were solved 
numerically, giving excellent agreement with direct simulations of the 
model. 

The most surprising result was the very fast increase of the average 
earth quake size as one approaches the conservative limit. Obviously this 
is a consequence of the non-locality of the interaction, since this 
implies that one can have extremely large earth quakes without having 
large effects locally. Nevertheless, avalanche size 
distributions decay exponentially for any non-zero dissipation.

Our findings support the view \cite{grass1} that scaling in the OFC 
model with inhomogeneous boundary conditions is due to a subtle 
interplay between partial synchronization and desynchronization. The 
inhomogeneity of the bc drives the synchronization in the bulk, 
building up large coherent patches, but occasionally the driving is too 
strong and the synchronization breaks down. Explicit observations of 
these patterns \cite{grass1,middleton} support this view. In a random 
neighbor model such structures cannot build up, of course, and the 
mechanism driving the system into a self-organized critical state is 
absent.

While we concentrated here on the OFC model, we showed that our methods 
can also be applied in other related models. In particular, we studied 
the Feder-Feder model in some detail. We showed that it also has no 
phase transition in the dissipative regime, and that the toppling rate 
and the mean avalanche size diverge in the conservative limit.

\vspace{.5cm}

Acknowledgement:\\
The authors want to thank H.\ Flyvberg for interesting discussions. 
This work was supported by the DFG within the Graduiertenkolleg
`Feldtheoretische und numerische Methoden in der Elementarteilchen-
und Statistischen Physik', and within Sonderforschungsbereich 237.

\eject

\end{document}